\documentclass[floats,twocolumn,prb]{revtex4}
\usepackage{graphicx}
\usepackage{dcolumn}
\usepackage{bm}
\usepackage{mathrsfs}

\begin{document}
\preprint{APS/123-QED}

\title{Electronic correlation-driven orbital polarization transitions in the orbital-selective Mott compound
         Ba$_2$CuO$_{4-\delta}$}

\author{Yu Ni$^1$}
\author{Ya-Min Quan$^2$}
\author{Jingyi Liu$^1$}
\author{Yun Song$^1$}\emph{}
\thanks{yunsong@bnu.edu.cn}
\author{Liang-Jian Zou$^2$}\emph{}
\thanks{zou@theory.issp.ac.cn}

\affiliation{%
$^1$Department of Physics, Beijing Normal University, Beijing 100875, China}%

\affiliation{%
$^2$Key Laboratory of Materials Physics, Institute of Solid State Physics,
 Chinese Academy of Sciences, P. O. Box 1129, Hefei 230031, China}%

\date{\today}

\begin{abstract}
  The electronic states near the Fermi level of recently discovered superconductor
  Ba$_2$CuO$_{4-\delta}$ consist primarily of the Cu $d_{x^2-y^2}$ and $d_{3z^2-r^2}$ orbitals.
  We investigate the electronic correlation effect and the orbital polarization of an effective
  two-orbital Hubbard model mimicking the low-energy physics of   Ba$_2$CuO$_{4-\delta}$ in the
  hole-rich regime by utilizing the dynamical mean-field theory with   the {\it Lanczos} method
  as the impurity solver.
  We find that the hole-overdoped Ba$_2$CuO$_{4-\delta}$ with $3d^8$ (Cu$^{3+}$) is in the
  orbital-selective Mott phase (OSMP) at half-filling, and the typical
  two-orbital feature  remains in Ba$_2$CuO$_{4-\delta}$ when the electron filling approaches
  $n_e\sim 2.5$, which  closely approximates to the experimental hole doping for the emergence
  of the high-$T_c$ superconductivity.
  We also obtain that the orbital polarization is very stable in the OSMP, and the multiorbital
  correlation can drive orbital polarization transitions. These results indicate that in
  hole-overdoped Ba$_2$CuO$_{4-\delta}$ the OSMP physics and orbital polarization,
  local magnetic moment and spin or orbital fluctuations still exist.
  We propose that our present results are also applicable to Sr$_2$CuO$_{4-\delta}$ and other
  two-orbital cuprates, demanding an unconventional multiorbital superconducting scenario in
  hole-overdoped high-$T_c$ cuprates.
\end{abstract}


\maketitle



\section{INTRODUCTION}

The involvement of two Cu 3$d$ orbitals in the superconducting (SC) states in the recently discovered
high-$T_c$ superconducting (HTSC) compound Ba$_{2}$CuO$_{4-\delta}$ with $T_c$=73 K,
\cite{JinCQ-PNAS-2019} as well as the early discovered compound Sr$_{2}$CuO$_{4-\delta}$ with
$T_c$=95 K, \cite{JinCQ-2006,Jin-2009,Han-1994}
greatly challenges the prevailing single-orbital scenario in conventional HTSC cuprates.
In the previous cuprates La$_{2}$CuO$_{4}$, and YBa$_{2}$Cu$_{3}$O$_{6}$, {\it etc.}, the parent
phases of these undoped compounds are charge transfer insulator or Mott insulator, where the active
Cu 3$d_{x^2-y^2}$ orbital is singly occupied and the ground state is {\it N\'eel}
antiferromagnetic insulator. Once holes are doped into the O 2$p_{x}$ and 2$p_{y}$ orbitals,
the strong O 2$p$-Cu 3$d$ hybridization and large charge transfer gap form the {\it Zhang-Rice}
singlet,\cite{ZR-singlet} and an effective single-orbital
$t-J$ model is proposed for describing the low-energy physics of doped cuprates.\cite{tJ-1989}
Such an effective single-orbital scenario addressed many experimental results,\cite{RMPcuprates-2010}
demonstrating the reasonability of the model.
The essential electronic states in Ba$_{2}$CuO$_{4-\delta}$ do not fall into this scenario: first,
neither Ba$_{2}$CuO$_{3}$ nor Ba$_{2}$CuO$_{4}$ is a charge transfer insulator,
instead, its charge transfer gap is rather small;
second, both the Cu $3d_{3z^2-r^2}$  and $3d_{x^2-y^2}$ orbitals appear near the
Fermi energy in SC Ba$_2$CuO$_{4-\delta}$. \cite{JinCQ-PNAS-2019}
This suggests that Ba$_{2}$CuO$_{4-\delta}$ is a multiorbital superconductor.
This scenario which is completely different from that of the well-known $t-J$ model brings about
the assumptions of two
SC dome phases and orbital selective superconductivity.\cite{Scalapino-2019,Ziqiang Wang-2018}

The detail inspections to the electronic properties of Ba$_{2}$CuO$_{4-\delta}$ will provide
new insight or even new scenario, especially the electronic states of SC Ba$_{2}$CuO$_{4-\delta}$
lie in Ba$_{2}$CuO$_{4}$ and in Ba$_{2}$CuO$_{3}$.
It is well-known that La$_2$CuO$_4$ and derivative cuprates are strongly correlated systems. We expect
that Ba$_{2}$CuO$_{4-\delta}$ is a correlated system, though it can be viewed as a hole-overdoped
compound. Recently, Liu $et$ $al$ proposed that Ba$_{2}$CuO$_{3}$ is an antiferromagnetic
insulator\cite{XiangT-2019}
and should be the parent phase; however, Maier $et$ $al$ suggested that Ba$_{2}$CuO$_{4}$ should be
the parent phase.\cite{Scalapino-2019}
To resolve such a dispute, it is crucial to clarify the role of the electronic correlation in
Ba$_{2}$CuO$_{4-\delta}$ in the hole-rich regime.
Correspondingly, one may also ask what the role of Hund's rule coupling plays in such a multiorbital
system and how the quantum phases evolve with increasing hole concentration.
\cite{Scalapino-2019,Ziqiang Wang-2018,XiangT-2019,ZhangFC-2020,LiYH-2020}
These issues are important since orbital and magnetic fluctuations
are closely related to the ground-state magnetism.

On the other hand, we notice that compared with antiferromagnetically insulating Ba$_{2}$CuO$_{3}$,
Ba$_{2}$CuO$_{4}$ exhibits a paramagnetically metallic ground state, \cite{XiangT-2019} though it is
stoichiometric 3$d^7$ configuration. How can such a paramagnetically metallic phase be stable
in an integer-filling correlated electron system?
At present it is not clear that what role the electronic correlation plays in the paramagnetically
metallic ground states of Ba$_{2}$CuO$_{4}$. Meanwhile, the orbital  polarization character of
the hole-overdoped Ba$_{2}$CuO$_{4-\delta}$,
which is essential for SC pairing symmetry, is also profoundly affected by the electronic correlation.
These facts urge us to clarify the role of electronic correlations in the two-orbital compound
Ba$_{2}$CuO$_{4-\delta}$, as well as Sr$_{2}$CuO$_{4-\delta}$.

In this paper we use the dynamical mean-field theory (DMFT) \cite{Georges-1996,Kotliar-2006,Rohringer-2018}
with the {\it Lanczos} method as its impurity solver to investigate the influences of Coulomb
correlation and hole doping on the electronic states in a two-orbital Hubbard model, which is
applicable for the compressed Ba$_{2}$CuO$_{4-\delta}$ compound.
Our results suggest that Ba$_2$CuO$_{4-\delta}$ compound has a typical two-orbital
character, even when the electron filling approaches $n_e\sim 2.5$, which is very close to the
optimal hole doping of the high-$T_c$ superconducting phase.\cite{JinCQ-PNAS-2019}
We demonstrate that the hole-overdoped Ba$_2$CuO$_{4-\delta}$ is in the orbital-selective
Mott phase (OSMP) at half-filling with two electrons in the two
$e_g$ orbitals, which is regarded as the the parent phase of Ba$_2$CuO$_{4-\delta}$ compound.
Our results also show that the orbital polarization is extremely stable in the OSMP region,
providing a direct evidence for the occurrence of the OSMP.
The orbital polarization transitions can be driven by the multiorbital correlations in the Ba$_2$CuO$_{4-\delta}$, especially in the hole-rich region.

This paper is organized as follows.
In Sec.~II we introduce a two-orbital Hubbard model for the hole-overdoped Ba$_2$CuO$_{4-\delta}$,
and we explain the numerical method adopted in our study: the DMFT approach with the {\it Lanczos}
solver.
In Sec.~III we demonstrate the effects of electronic correlation and hole doping on the electronic
states by analyzing the phase diagrams of the hole-overdoped Ba$_2$CuO$_{4-\delta}$.
The principal findings of this paper are summarized in Sec.~IV.

\section{Model and method}

On account of the crossing to the Fermi energy for both the two bands formed
from the Cu $3d_{x^2-y^2}$ and $3d_{3z^2-r^2}$ orbitals of the compressed
Ba$_2$CuO$_{4-\delta}$ compound, we investigate the electronic states described
by a two-orbital Hamiltonian $H=H_t+H_I$, where the tight-binding (TB) Hamiltonian
$H_t$ reads\cite{Scalapino-2019}
\begin{eqnarray}
    H_t&=&-\sum_{ij}\sum_{l \sigma}
    t^{(ij)}_{ll} d^{\dag}_{il\sigma}d_{jl\sigma}
    -\sum_{ij}\sum_{l\neq l', \sigma}
    t^{(ij)}_{ll'}d^{\dag}_{il\sigma}d_{jl'\sigma}
    \nonumber\\
    &&+\sum_{il\sigma}(\epsilon_l-\mu) d^{\dag}_{il\sigma}d_{il\sigma}.
\end{eqnarray}
$d^{\dag}_{il\sigma}$ ($d_{il\sigma}$) is an electron creation (annihilation)
operator for orbital $l$ (=1 for $d_{x^2-y^2}$ and 2 for $d_{3z^2-r^2}$) at site $i$ with spin $\sigma$.
$t^{(ij)}_{11/22}$ and $t^{(ij)}_{12}$ represent the intraorbital and interorbital hoppings between sites
$i$ and $j$, respectively. $\epsilon_l$ represents the on-site energy of
orbital $l$, and the crystal-field splitting is given as
$\epsilon_d=\epsilon_1-\epsilon_2$. $\mu$ denotes the chemical potential.

The interaction Hamiltonian $H_I$ is exactly the same as the correlation part of the standard two-orbital
Hubbard model,\cite{oles-1983,Koga-2005}
\begin{eqnarray}
    H_I&=&\frac{U}{2}\sum_{il\sigma}n_{il\sigma}n_{il\bar{\sigma}}+\sum_{i,l<l',\sigma\sigma'}
    (U'-\delta_{\sigma\sigma'}J_H)n_{il\sigma}n_{il'\sigma'}
    \nonumber\\
    &&+\frac{J_H}{2}\sum_{i,l\neq l',\sigma} d^{\dag}_{il\sigma}d^{\dag}_{il\bar{\sigma}}d_{il'\bar{\sigma}}d_{il'\sigma}
    \nonumber\\
    &&+\frac{J_H}{2}\sum_{i,l\neq l',\sigma\sigma'} d^{\dag}_{il\sigma}d^{\dag}_{il'\sigma'}d_{il\sigma'}d_{il'\sigma},
\label{Eq:TOHub}
\end{eqnarray}
where $U$ ($U'$) corresponds to the intraorbital (interorbital) interaction,
and $J_H$ is the Hund's rule coupling. For the systems with spin rotation symmetry,
we have $U=U'+2J_H$.

Ba$_2$CuO$_{4-\delta}$ can be viewed as a hole-overdoped compound.
The nominal 2(1-$\delta$) holes per Cu are doped in the Ba$_2$CuO$_{4-\delta}$
compound,\cite{JinCQ-PNAS-2019} giving a relation between $\delta$ and the hole
concentration $x_h$ as $\delta=1-x_h/2$.
Accordingly, the copper valence can be expressed as Cu$^{2+x_h}$ for a hole concentration
$x_h$.\cite{Scalapino-2019}
At half-filling  with $\delta=0.5$, there are two electrons in the $d_{x^2-y^2}$
and $d_{3z^2-r^2}$ orbitals of Cu$^{3+}$  in Ba$_2$CuO$_{4-\delta}$.

Based on the DFT-calculated band structures of the compressed
Ba$_2$CuO$_{4-\delta}$ compound with hole concentration $x_h=1$,\cite{Scalapino-2019}
the model parameters of the TB Hamiltonian $H_t$ in Eq.~(1) are given
in Table I, including the hopping parameters for the $1_{\rm st}$ ($t$),
$2_{\rm nd}$ ($t'$), and $3_{\rm rd}$ ($t''$) nearest neighbors, as well as the
on-site energy $\epsilon_l$. Through the Fourier transformation, $H_t$ is changed to
\begin{eqnarray}
H_{0}(k)=\sum_{k \sigma} \sum_{l l^{\prime}}\left\{\xi_{l l^{\prime}}(k)+\left[\epsilon_{l}-\mu\right] \delta_{l l^{\prime}}\right\} d_{l \sigma}^{\dagger}(k) d_{l^{\prime} \sigma}(k),
\label{Eq:H0}
\end{eqnarray}
with
\begin{eqnarray}
\xi_{11 / 22}(k) &=&2 t_{11 / 22}\left(\cos k_{x}+\cos k_{y}\right)\nonumber\\
&&+4 t_{11 / 22}^{\prime} \cos k_{x} \cos k_{y}\nonumber\\
&&+2 t_{11 / 22}^{\prime \prime}\left(\cos 2 k_{x}+\cos 2 k_{y}\right),
\end{eqnarray}
and
\begin{eqnarray}
\xi_{12}(k)&=&\xi_{21}(k)\nonumber\\
&=&2 t_{12}\left(\cos k_{x}-\cos k_{y}\right)\nonumber\\
&&+2 t_{12}^{\prime \prime}\left(\cos 2 k_{x}-\cos 2 k_{y}\right).
\end{eqnarray}

The energy of the $d_{3z^2-r^2}$ orbital varies as $\epsilon_2=0.661-5(1-x_h)$
with decreasing hole concentration $x_h$, while the energy $\epsilon_1$ of the orbital $d_{x^2-y^2}$
is kept as a constant.\cite{Scalapino-2019}
For the conditions with $x_h<1$, the chemical potential $\mu$ has been adjusted to keep
an electron filling $n_e=1+2\delta$ ($n_e=3-x_h$ ) for the hole-overdoped Ba$_2$CuO$_{4-\delta}$.

\begin{table}
	\caption{Model parameters of the TB Hamiltonian
    of Ba$_2$CuO$_{4-\delta}$ at half-filling in eV.\cite{Scalapino-2019}
		\label{tab:tab1}}
	\begin{tabular}{|c|c|c|c|c|}
		\hline
		 & on-site  & 1$_{\rm st}$ hop- & 2$_{\rm nd}$ hop- & 3$_
		 {\rm rd}$ hop-\\
         &  energy ($\epsilon$) &  ping ($t$) & ping  ($t^\prime$) & ping ($t^{\prime\prime}$)\\	\hline
		orbital $d_{x^2-y^2}$ & -0.222 & 0.504
		 & -0.067 & 0.130 \\	\hline
		orbital $d_{3z^2-r^2}$ & 0.661 & 0.196 & 0.026 &
		0.029 \\	\hline
		inter-orbital & 0 & -0.302 & 0 & -0.051
		\\	\hline 
	\end{tabular}
\end{table}

We map the lattice Hamiltonian on to an impurity
model with fewer degrees of freedom,
\begin{eqnarray}
{H_{imp}} &=& \sum\limits_{ml\sigma} {{\varepsilon_{ml\sigma}}} c_{ml\sigma}^\dag {c_{ml\sigma}}+ \mathop \sum \limits_{l\sigma} (\epsilon_l-\mu)d_{l\sigma}^\dag {d_{l\sigma}} \nonumber\\
&&+ \sum\limits_{ll'm\sigma} {{V_{ll'm\sigma}}\left( {d_{l\sigma}^\dag {c_{ml'\sigma}} + c_{ml'\sigma}^\dag {d_{l\sigma}}} \right)}  \nonumber\\
&& + {H_I^{imp}}\,
\label{Eq:IMP}
\end{eqnarray}
with
\begin{eqnarray}
    H_I^{imp}&=&\frac{U}{2}\sum_{l\sigma}n_{l\sigma}n_{l\bar{\sigma}}+\sum_{l<l',\sigma\sigma'}
    (U'-\delta_{\sigma\sigma'}J_H)n_{l\sigma}n_{l'\sigma'}
    \nonumber\\
    &&+\frac{J_H}{2}\sum_{l\neq l',\sigma} d^{\dag}_{l\sigma}d^{\dag}_{l\bar{\sigma}}d_{l'\bar{\sigma}}d_{l'\sigma}
    \nonumber\\
    &&+\frac{J_H}{2}\sum_{l\neq l',\sigma\sigma'} d^{\dag}_{l\sigma}d^{\dag}_{l'\sigma'}d_{l\sigma'}d_{l'\sigma},
\end{eqnarray}
where $c^{\dag}_{ml\sigma}$ ($c_{ml\sigma}$) denotes the creation (annihilation) operator for
the bath lattice of orbital $l$, $\varepsilon_{ml\sigma}$ denotes the energy of
the $m$-th environmental bath of orbital $l$,
and $V_{ll'm\sigma}$ represents the coupling
between the orbital $l$ of the impurity site and environmental bath of orbital $l'$.

The Green's function of the two-orbital impurity model can be expressed as
a $2\times 2$ matrix,
\begin{eqnarray}
{\bm{G}_{imp}}\left( {i{\omega _n}} \right) = \left( {\begin{array}{*{20}{c}}
{{G_{11}}\left( {i{\omega _n}} \right)}&{{G_{12}}\left( {i{\omega _n}} \right)}\\
{{G_{21}}\left( {i{\omega _n}} \right)}&{{G_{22}}\left( {i{\omega _n}} \right)}
\end{array}} \right).\
\label{Eq:G_IMP}
\end{eqnarray}
The Green's function $\bm{G}_{imp}$ at zero-temperature is calculate by the {\it Lanczos} solver
\cite{Dagotto-1994,Caffarel-1994,Capone-2007}.
We choose a bath size $n_b=3$ in our calculations.
It has been proved that the critical values of the OSMT in a two-orbital Hubbard
model are almost the same when the bath size is taken as $n_b\geq 3$ in the
DMFT calculations with {\it Lanczos} solver.\cite{NiuYK-2019}

In the {\it Lanczos} procedure,\cite{Dagotto-1994} the diagonal matrix elements of
the Green's function $G_{ll}$ are expressed as
\begin{eqnarray}
G_{ll}\left( \omega\right)=G_{ll}^{(+)}\left( \omega\right)+G_{ll}^{(-)}\left( \omega\right),
\end{eqnarray}
where
\begin{eqnarray}
G_{ll}^{(+)}\left( \omega\right)=\frac{\left\langle\phi_{0}\left|d_l d_l^{\dagger}\right| \phi_{0}\right\rangle}{ \omega-a_{0}^{(+)}-\frac{b_{1}^{(+) 2}}{ \omega-a_{1}^{(+)}-\frac{b_{2}^{(+) 2}}{ \omega-a_{2}^{(+)}-\ldots}}},
\label{G-positive}
\end{eqnarray}
and
\begin{eqnarray}
G_{ll}^{(-)}\left( \omega\right)=\frac{\left\langle\phi_{0}\left|d_l^{\dagger} d_l\right| \phi_{0}\right\rangle}{ \omega+a_{0}^{(-)}-\frac{b_{1}^{(-) 2}}{ \omega+a_{1}^{(-)}-\frac{b_{2}^{(-) 2}}{ \omega+a_{2}^{(-)}-\ldots}}}.
\label{G-negative}
\end{eqnarray}
$\left|\phi_{0}\right\rangle$ is the ground state. $a_{n}^{(+)}$ and $b_{n}^{(+)}$ are the
elements of tridiagonal form of the Hamiltonian matrix, constructed from the initial state $d_{l}^{\dagger}\left|\phi_{0}\right\rangle / \sqrt{\left\langle\phi_{0}\left|d_{l} d_{l}^{\dagger}\right| \phi_{0}\right\rangle}$, and $a_{n}^{(-)}$ and $b_{n}^{(-)}$
are correspondingly obtained by another initial state
$d_{l}\left|\phi_{0}\right\rangle / \sqrt{\left\langle\phi_{0}\left|d_{l}^{\dagger} d_{l}\right| \phi_{0}\right\rangle}$.
The off-diagonal elements $G_{12} $ and $G_{21} $ can be also obtained by the {\it Lanczos} method,
based on the relation
\begin{eqnarray}
G_{12}^{}\left( {i{\omega _n}} \right) &=& G_{21}^{}\left( {i{\omega _n}} \right) \nonumber\\
& =&\frac{1}{2}\left[ {G_{1+2,1+2}^{}\left( {i{\omega _n}} \right) - \sum\limits_l {G_{ll}^{}\left( {i{\omega _n}} \right)} } \right],\
\end{eqnarray}
where $G_{1+2,1+2}$ is a combined Green's function,
$G_{1+2,1+2}=\langle\langle d_{1} + d_{2} | d_{1}^\dag  + d_{2}^\dag \rangle\rangle$,
which is calculated in the {\it Lanczos} method by replacing the operators $d_l^{\dag}$ and $d_l$ as $d_1^{\dag}+d_2^{\dag}$ and $d_1+d_2$, respectively.

The Weiss function of the impurity model can be obtained through the parameters of the impurity
Hamiltonian by
\begin{eqnarray}
\bm{{\cal G}}_0^{imp}{\left( {i{\omega _n}} \right)^{ - 1}} =( i{\omega _n} + \mu  )\bm{I}- \bm{\Delta} (i{\omega _n}),\
\end{eqnarray}
where $\bm{{\cal G}}_0^{imp}{\left( {i{\omega _n}} \right)}$ and $\bm{\Delta} (i{\omega _n})$ are
$2\times 2$ matrices, and symbol $\bm{I}$ denotes the identity matrix. The hybridization matrix
$\bm{\Delta} (i{\omega _n})$ is defined as
\begin{eqnarray}
\bm{\Delta} \left( {i{\omega _n}} \right) = \left( {\begin{array}{*{20}{c}}
  {{\Delta _{11}}\left( {i{\omega _n}} \right)}&{{\Delta _{12}}\left( {i{\omega _n}} \right)} \\
  {{\Delta _{21}}\left( {i{\omega _n}} \right)}&{{\Delta _{22}}\left( {i{\omega _n}} \right)}
\end{array}} \right),\
\end{eqnarray}
with
\begin{eqnarray}
{\Delta _{{l_1}{l_2}}}\left( {i\omega _n} \right) \equiv \sum\limits_{ml} {\frac{{{V_{{l_1}lm}}{V_{{l_2}lm}}}}{{i\omega_n  - {\varepsilon_{ml}}}}}. \
\end{eqnarray}
We calculate the $2\times 2$ self-energy matrix ${\bm{\Sigma} _{imp}}(i\omega _n)$ of the impurity Hamiltonian by the Dyson equation
\begin{eqnarray}
{\bm{\Sigma} _{imp}}(i\omega _n) = \bm{{\cal G}}_0^{imp}{\left( {i{\omega _n}} \right)^{ - 1}}-{\bm{G}_{imp}}{\left( {i{\omega _n}} \right)^{ - 1}},\
\end{eqnarray}
and the $2\times 2$ lattice Green's function matrix ${\bm{G}_{lat}}\left( {i\omega _n} \right)$ is obtained by
\begin{eqnarray}
{\bm{G}_{lat}}\left( {i\omega _n} \right) &=& \frac{1}{N}\sum\limits_k {\bm{G}\left( {i\omega _n,k} \right)}  \nonumber\\
&=& \frac{1}{N}\sum\limits_k {\frac{1}{{i\omega _n{\bm I} - {\bm{H}_0}\left( k \right) - {\bm{\Sigma} _{imp}}(i\omega _n)}}},\nonumber\\ \
\end{eqnarray}
where ${\bm{H}_0}$ is the matrix representation of Eq.~(\ref{Eq:H0}).

We build the DMFT self-consistent loop with
${\bm G}_{imp}(i \omega _n)={\bm G}_{lat}(i\omega _n)$ to determine the parameters
$\varepsilon_{ml}$ and $V_{ll'm}$.
Analytic continuation is also performed to obtain the real frequency
Green's function ${\bm G(\omega)}$.\cite{Georges-1996}

\begin{figure}[htbp]
\includegraphics[scale=0.70]{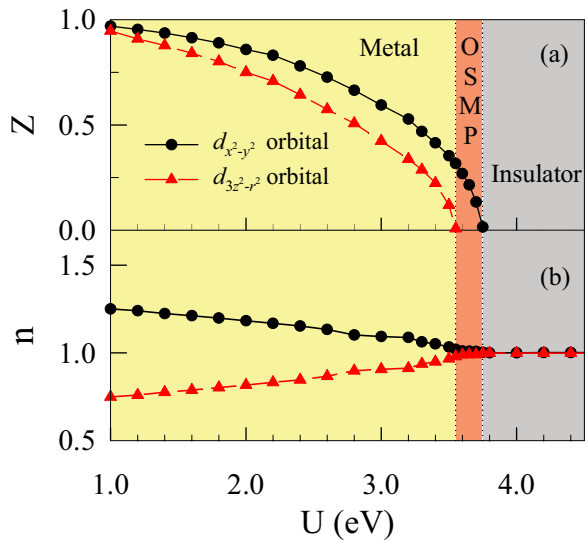}
\caption{(Color online)
  Orbital-resolved quasiparticle weight $Z_l$ (a) and electron occupation $n_l$ (b) as
  a function of interaction $U$ when $J_H=0.25U$ for Ba$_2$CuO$_{4-\delta}$ at half-filling.
  Between the metallic (yellow) and insulating (gray) phases, an OSMP (orange)
  occurs in a narrow interaction region with 3.55~eV $\leq U <$~3.75 eV.
  }
\label{fig:NZU}
\end{figure}

We calculate the orbital-resolved spectral density of each orbital by
\begin{equation}
A_{l}(\omega)=-\frac{1}{\pi}\rm{Im}\it{G^{(lat)}_{ll}(\omega+i\eta)},
\end{equation}
where $\eta$ is an energy broadening factor.
Then, the orbital projected optical conductivity can be expressed approximately as
\begin{eqnarray}
\sigma_{l}(\omega)&=&\pi\int_{-\infty}^{\infty}d\epsilon D_{l}(\epsilon)
     \int_{-\infty}^{\infty}\frac{d\omega^{\prime}}
     {2\pi}A_{l}(\omega^{\prime})A_{l}(\omega^{\prime}+\omega)
     \nonumber\\
  &&\times\frac{n_{f}^{(l)}(\omega^{\prime})-n_{f}^{(l)}(\omega^{\prime}+\omega)}{\omega},
\label{optical-cond}
\end{eqnarray}
where $n_{f}(\omega)$ is the Fermi function, and
$D_l$ represents the density of states (DOS) of the TB Hamiltonian.
We neglect the vertex correction to the current operator, and the off-diagonal
elements are also neglected in our calculations.

To explore possible orbital polarization and orbital ordering in Ba$_2$CuO$_{4-\delta}$,
we also calculate the local orbital squared moment $\left\langle T_{z}^{2}\right\rangle$ of the
two orbitals by\cite{Zou-2008}
\begin{equation}
\left\langle T_{z}^{2}\right\rangle=\langle (\hat{n}_{l_1}-\hat{n}_{l_2})^2 \rangle,
\end{equation}
from which we could obtain the evolution of orbital polarization with increasing electronic correlation.

\section{Results and Discussions}

We study the electronic states in Ba$_2$CuO$_{4-\delta}$ as the hole doping varies from
intermediate doping region to highly overdoped region, and we also pay close attention to the optimal
hole-doping region around the electron filling $n_e\sim 2.5$, where the high-$T_c$
superconductivity occurs in Ba$_2$CuO$_{4-\delta}$.
Firstly, we study the orbital selective Mott transition (OSMT) in the strongly overdoped
system for $3d^8$ (Cu$^{3+}$) with two electrons in the two $e_g$ orbitals, i.e.,
at half-filling with $n_e=2$, which is regarded as the the parent phase of
Ba$_2$CuO$_{4-\delta}$ compound.

\begin{figure}[htbp]
\includegraphics[scale=0.45]{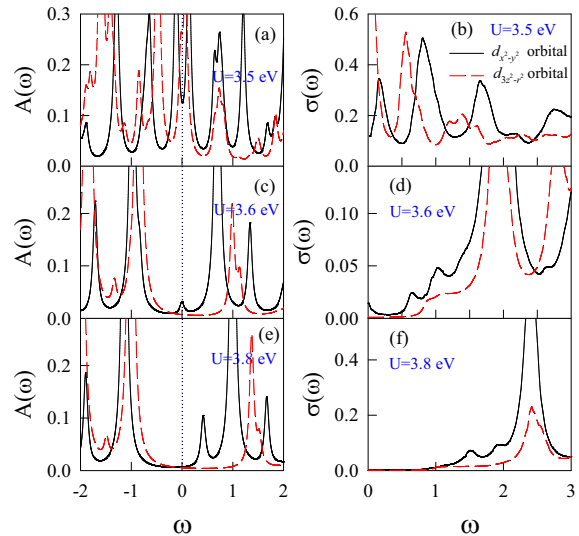}
\caption{(Color online)
  Effect of interaction $U$ on the orbital-resolved spectral
  density $A_l(\omega)$ (left panel) and the corresponding orbital-dependent optical
  conductivity $\sigma_l(\omega)$ (right panel) of the two bands
  of Ba$_2$CuO$_{4-\delta}$ with $J_H=0.25U$ at half-filling. The Fermi energy is
  denoted by a blue dotted line, and the energy broadening
  is given as $\eta$=0.05~eV.
  }
\label{fig:DOSOC-J4}
\end{figure}

\subsection{OSMT at half-filling}

Half filled Ba$_2$CuO$_{4-\delta}$ with $n_e=2$ displays prominent OSMP character under the correlated effect.
In Fig.~\ref{fig:NZU}(a) we present the orbital-dependent quasiparticle weight $Z_l$, $Z_l=(1-\frac{\partial}{\partial\omega}\rm{Re}
\it{\Sigma_l(\omega)|_{\omega=0}})^{-1}$, \cite{Liebsch-2007}
as a function of the intraorbital interaction $U$ when $J_H=0.25U$ at half-filling.
When $U<3.55$~eV, the two-band system is metallic with finite quasiparticle weights
of the two orbitals. By contrast, the insulating phase with zero $Z_{l}$
is stable for both orbitals when $U>3.75$~eV.
An OSMP occurs between the metallic and insulating phases with $3.55~$eV$\leq U<3.75~$eV, in
which the narrow $d_{3z^2-r^2}$ band behaves insulating, while the wide $d_{x^2-y^2}$
band is still metallic.
The $U$-dependence of the electron occupations $n_l$ are shown in Fig.~\ref{fig:NZU}(b).
We find that the electrons transfer from the wide $d_{x^2-y^2}$ band to the the narrow
$d_{3z^2-r^2}$ band with increasing $U$ at half-filling.
Both bands become singly occupied ($n_l=1$) after the OSMT, indicating that the two electrons
in the two Cu $e_g$ orbitals distribute uniformly in both the OSMP and insulating phase,
as observed in other degenerate two-orbital systems.
\cite{Dongen-2009,Koga-2004,Anisimov-2002,YKNiu-2018}

\begin{figure}[htbp]
\includegraphics[scale=0.65]{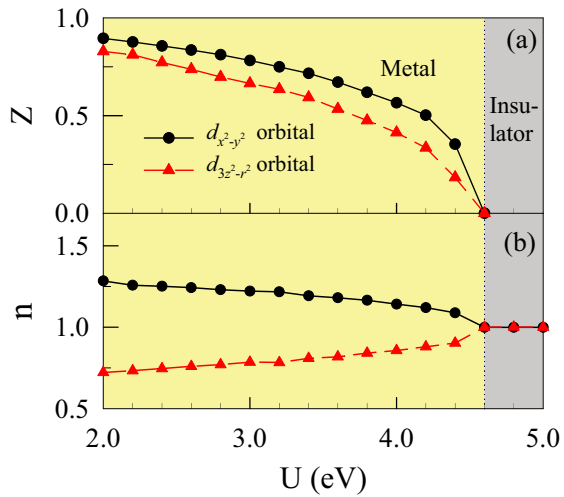}
\caption{(Color online)
  Quasiparticle weight (a) and electron occupation (b) as a function of interaction $U$
  when $J_H=0.125U$ at half-filling. A direct transition from metallic phase to insulating
  phase is found.
  }
\label{fig:MIT-J8}
\end{figure}

Fig.~\ref{fig:DOSOC-J4} shows the orbital-resolved spectrum  $A_l(\omega)$ and
optical conductivity $\sigma_l(\omega)$  obtained in different phases.
In the metallic phase with $U=3.5$~eV, both bands have a finite spectral weight at the
Fermi level, and there is a Drude peak in the optical conductivity accordingly,
as shown in Fig.~\ref{fig:DOSOC-J4}(a) and \ref{fig:DOSOC-J4}(b) respectively.
At $U$=3.6~eV a small resonance peak can be found in the DOS of
the metallic wide $d_{x^2-y^2}$ band, accompanied with a small Drude peak in its
optical conductivity.
Meanwhile, a Mott gap opens around the Fermi level in the narrow $d_{3z^2-r^2}$ band, and
the Drude weight is zero for its optical conductivity, demonstrating the
well-defined OSMP character.\cite{JSun-2015,Costi-2007,Medici-2011}
Insulating phase appears at $U$=3.8~eV, where a Mott gap opens in the DOS and
there is no Drude peak in the optical conductivity for both bands as shown in
Fig.~\ref{fig:DOSOC-J4}(e) and \ref{fig:DOSOC-J4}(f), respectively.

Decreasing the Hund's rule coupling to $J_H=0.125U$, the Mott transitions in the
two bands occur simultaneously at $U=U_{c1}=U_{c2}=4.6$~eV,
as shown in Fig.~\ref{fig:MIT-J8}(a).
This finding is in agreement with the early result that a large $J_H$ promotes the
OSMT at half-filling by strongly suppressing the coherence scale to block the orbital
fluctuations.\cite{Georges-2013,Liebsch-2005,deMedici-2011,JSun-2015}
Fig.~\ref{fig:MIT-J8}(b) shows that both bands become half-filled  after the Mott
transition.

\begin{figure}[htbp]
\includegraphics[scale=0.55]{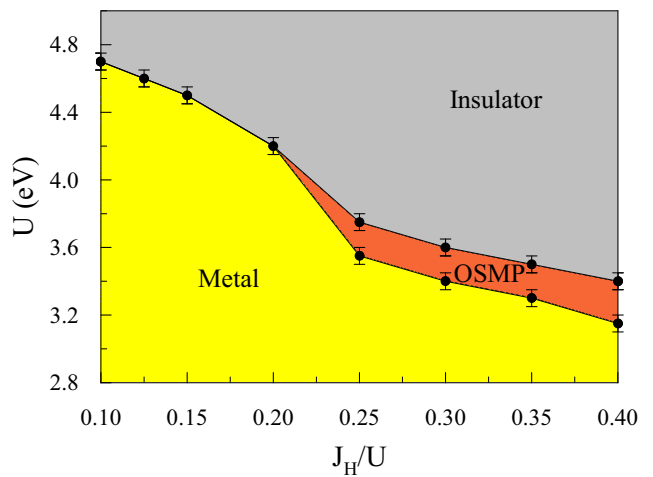}
\caption{(Color online)
  Phase diagram of Ba$_2$CuO$_{4-\delta}$ at half-filling.
  The critical values of Mott transitions in the narrow and wide bands decrease with
  increasing $J_H/U$, leading the OSMP region to be wider accordingly.
  The OSMT vanishes if $J_H/U$ is smaller than 0.2.
  }
  \label{fig:JU-HF}
\end{figure}

We construct the $U$-$J_H$ phase diagram in Fig.~\ref{fig:JU-HF}. One observes that there
exists a narrow region of the OSMP between the weakly-correlated metallic phase and strongly-correlated
Mott insulating phase when $J_H>0.2U$, which is getting to broaden with increasing $J_H$.
It is obvious that large $J_H$ is beneficial to the occurrence of OSMP. The OSMP vanishes
when $J_H<0.2U$ because Coulomb correlation and Hund's rule coupling are inferior to
the crystal-field splitting, in agreement with the previous
results.\cite{Liebsch-2005,deMedici-2011,Georges-2013,Jakobi-2013}
%
%
In the region of $J_H>0.2U$, the system undergoes the transitions of a metallic phase to an OSMP
and of an OSMP to an insulating phase as $U$ increases.
Because Ba$_2$CuO$_{4-\delta}$ at half-filling is at least an intermediate correlated
system,\cite{Scalapino-2019}
it should be an OSMP compound, or near the edge of the OSMP.

\begin{figure}[htbp]
\includegraphics[scale=0.50]{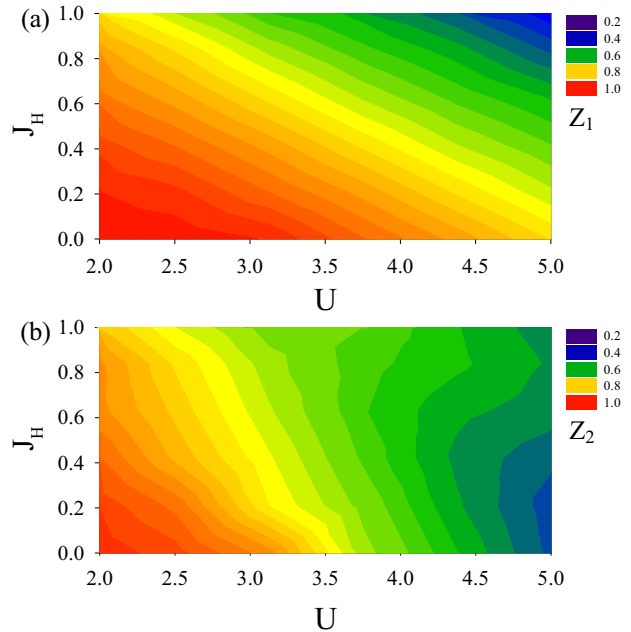}
\caption{(Color online)
  Effects of Hund's rule coupling $J_H$ and interaction $U$ on quasiparticle weight $Z_l$ of the
  wide band $d_{x^2-y^2}$ (upper panel) and the narrow band $d_{3z^2-r^2}$ (lower panel)
  for Ba$_2$CuO$_{4-\delta}$ when $n_e\sim 2.5$.
  }
\label{fig:ZD-x}
\end{figure}

There have been two possibilities regarding the parent compound of the SC Ba$_2$CuO$_{4-\delta}$.
One candidate is Ba$_2$CuO$_{4}$,\cite{Scalapino-2019} and the other one is
Ba$_2$CuO$_{3}$.\cite{XiangT-2019}
Our study suggests an alternative possibility that the half-filled Ba$_2$CuO$_{3.5}$ is the parent
compound. Increasing the electron filling by removing some oxygens from Ba$_2$CuO$_{3.5}$,
high-$T_c$ superconductivity emerges in Ba$_2$CuO$_{4-\delta}$
when $n_e\sim 2.5$.\cite{JinCQ-PNAS-2019}

\begin{figure}[htbp]
\includegraphics[scale=0.50]{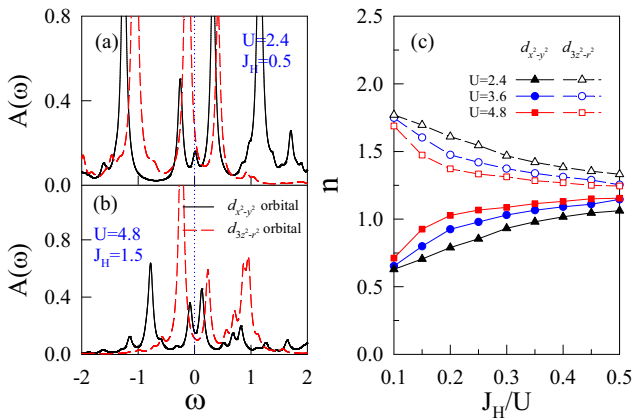}
\caption{(Color online)
  Orbital-resolved spectral density $A_l(\omega)$ of Ba$_2$CuO$_{4-\delta}$ with $n_e\sim 2.5$
  for different interactions: (a) $U$=2.4~eV and $J_H$=0.5~eV 
  and (b) $U$=4.8~eV and $J_H$=1.5~eV. The energy broadening is $\eta=0.05$~eV.
  (c) Orbital-resolved electron occupation $n_l$ as a function of  $J_H/U$ for different interactions:
  $U$=2.4~eV, 3.6~eV and 4.8~eV when the electron filling $n_e$ is around 2.5.
  }
\label{fig:SN-J}
\end{figure}

\subsection{Hole overdoping effect}

Focussing on the optimal hole doping for the occurrence of the high-$T_c$ superconductivity,
in Fig.~\ref{fig:ZD-x} we present the evolution of the orbital-resolved quasiparticle
weight $Z_l$ with increasing multiorbital interactions $U$ and $J_H$ in
Ba$_2$CuO$_{4-\delta}$  when the electron filling $n_e$ is around $2.5$.
Although the two bands are both good metal with large $Z_l$ in the weakly correlated region,
a significant difference between the quasiparticle weight distributions for the two bands can be
found in the strongly correlated region. It is obvious that the decline of the quasiparticle
weight of the narrow $d_{3z^2-r^2}$ band is mainly driven by the intraorbital interaction $U$.
On the other hand, a bad metallic character appears in the wide $d_{x^2-y^2}$ band when the
Hund's rule coupling is strong enough.
The two orbitals of Ba$_2$CuO$_{4-\delta}$ display different correlation features
in the strongly correlated region when $n_e\sim 2.5$.

We present the spectral density $A_l(\omega)$ and electron occupation $n_l$ at different $J_H/U$
in Fig.~\ref{fig:SN-J}.
In Fig.~\ref{fig:SN-J} (a) both bands have finite spectral weight at Fermi level
and the system displays prominent metal character when $U$=2.4~eV and $J_H$=0.5~eV.
With increasing correlations, we find two soft gaps in the DOS of both orbitals when
$U$=4.8~eV and $J_H$=1.5~eV shown in Fig.~\ref{fig:SN-J}(b), indicating that the
system becomes a bad metal in the strongly correlated region.
We show the orbital-dependent electron occupation in Fig.~\ref{fig:SN-J}(c) when the
electron filling is around 2.5.
Different from the results of half-filled systems shown in Fig.~\ref{fig:NZU}(b) and
Fig.~\ref{fig:MIT-J8}(b), electrons prefer to occupy the narrow $d_{3z^2-r^2}$ band.
It is worth noticing that both $J_H$ and $U$ tend to uniformly distribute electrons
within the two orbitals, and a finite Hund's rule coupling can make the wide
$d_{x^2-y^2}$ band to be around half-filled, indicating that both orbitals have significant
contributions. Therefore, our results give strong evidences of the two-orbital character
in Ba$_2$CuO$_{4-\delta}$ when $n_e\sim 2.5$, which corresponds to the experimental
hole-doping concentration for the occurrence of the high-$T_c$ superconductivity.

\begin{figure}[htbp]
\includegraphics[scale=0.50]{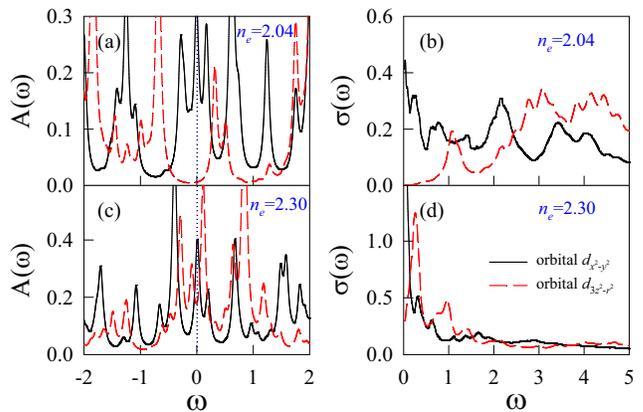}
\caption{(Color online)
  The orbital-resolved spectral density
  $A_l(\omega)$ (left panel) and the corresponding orbital-dependent optical conductivity
  $\sigma_l(\omega)$ (right panel) on different electron filling $n_e$ of Ba$_2$CuO$_{4-\delta}$
  when $J_H=0.25U$ and $U=3.6$~eV.
  The energy broadening is $\eta=0.05$~eV.
  }
\label{fig:DOSOC-x}
\end{figure}

To find out the influence of the hole doping in Ba$_2$CuO$_{4-\delta}$,
we extend our study to a wide doping region with $2.0 \leq n_e\leq 2.5$.
In Fig.~\ref{fig:DOSOC-x} we present the spectral density $A_l(\omega)$
and optical conductivity $\sigma_l(\omega)$
for different electron filling $n_e$ when $J_H=0.25U$ and $U=3.6$~eV,
where the system should be in an OSMP at half-filling ($n_e=2$)
based on the phase diagram shown in Fig.~\ref{fig:JU-HF}.
When we change the electron filling to $n_e=2.04$,
a Mott gap still opens in the narrow $d_{3z^2-r^2}$ band and
its optical conductivity is zero at $\omega=0$ correspondingly,
as shown in Fig.~\ref{fig:DOSOC-x}(a) and \ref{fig:DOSOC-x}(b).
Also, the wide $d_{x^2-y^2}$ band has a finite resonance peak and a large
Drude peak, indicating that the wide-band is in a metallic phase.
This indicates that an OSMP also occurs near half-filling.
When the electron filling is changed to $n_e=2.3$, the finite spectral
weights at Fermi level and the large Drude peaks for both bands indicate that
Ba$_2$CuO$_{4-\delta}$ transfers to a metallic phase,
as shown in Fig.~\ref{fig:DOSOC-x}(c) and \ref{fig:DOSOC-x}(d).

\begin{figure}[htbp]
\includegraphics[scale=0.50]{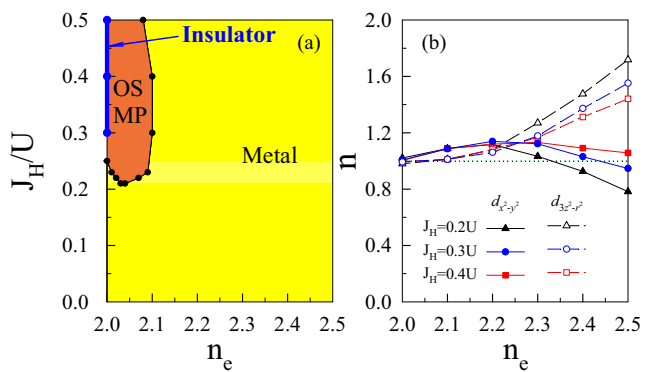}
  \caption{(Color online)
  (a) Phase diagram of Ba$_2$CuO$_{4-\delta}$ from the intermediate hole-doping region
  to the overdoped region with $2.0 \leq n_e <2.5$ when $U$=3.6~eV.
  (b) Orbital-resolved electron occupation as a function of $n_e$ for different $J_H$.
  The green dotted line indicates the half-filling of an orbital with $n=1$.
  The OSMP occurs in a hole-rich regime $2.0\leq n_e \leq 2.1$ for a not weak $J_H$,
  but the Mott insulating phase can only occur at half-filling with $n_e=2.0$ when the
  Hund's rule coupling is larger as $J_H>0.3U$.
  }
\label{fig:phase-x}
\end{figure}

In the phase diagram shown in Fig.~\ref{fig:phase-x}(a), an OSMP is found in
a highly overdoped region near half-filling,
which looks like an OSMP peninsula in the sea of metallic phase.
If the electron filling $n_e$ is more than 2.1, Ba$_2$CuO$_{4-\delta}$ with
$U$=3.6~eV can only be a metal no matter how large the Hund's rule coupling is.
On the other hand, the Mott insulating phase occurs only at half-filling with a
strong Hund's rule coupling as $J> 0.3U$. The critical value $J^c_H$ for the
OSMT takes the minimum value at $n_e \approx 2.04$.
Besides, the disappearance of the OSMP when $J_H\leq 0.2U$ provides further evidence
that the Hund's rule coupling can promote the OSMT,\cite{deMedici-2009} even away from
half-filling. Although the difference between the electron occupancies of the two
orbitals increases with increasing $n_e$, as shown in Fig.~\ref{fig:phase-x}(b),
the wide $d_{x^2-y^2}$ band is still approximately half-filled with $n_1\approx 0.95$
when $n_e=2.5$ and $J_H=0.3U$.
Our calculations demonstrate that Ba$_2$CuO$_{4-\delta}$ compound displays a typical two-orbital
character from the intermediate hole-doping region to the overdoped region, including the optimal doping
$n_e\sim 2.5$ for the occurrence of the high-$T_c$ superconductivity.
In the next subsection, we detect the correlation driven orbital polarization transitions in
Ba$_2$CuO$_{4-\delta}$, resulting from the electron transfer between the two orbitals.

\begin{figure}[htbp]
\includegraphics[scale=0.50]{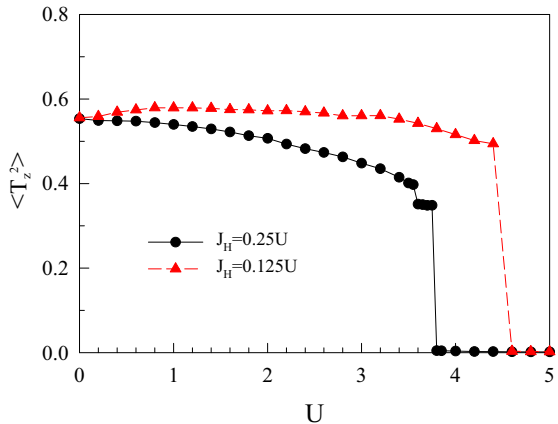}
\caption{(Color online)
  $U$ dependence of the squared moment $\left\langle T_{z}^{2}\right\rangle$
  for different Hund's rule coupling $J_H=0.25U$ (black line) and $J_H=0.125U$ (red line)
  at half-filling.
   }
\label{fig:Tz-HF}
\end{figure}

\subsection{Orbital Polarization}

In Fig.~\ref{fig:Tz-HF} we present the local orbital squared moment $\left\langle T_{z}^{2}\right\rangle$
as a function of the intraorbital interaction $U$ at half-filling. The model Hamiltonian employed in our
calculations is orbitally asymmetric, in which a large orbital polarization can exist in the metallic phase.
Based on the obtained phase diagram shown in Fig.~\ref{fig:JU-HF}, we find that the orbital squared
moments are finite when the system is metallic or in the OSMP, whereas a zero squared moment
$\left\langle T_{z}^{2}\right\rangle$=0 corresponds to the insulating phase. As a result, the $U$
dependence of the squared moment $\left\langle T_{z}^{2}\right\rangle$ displays a  stair-step profile
with a quickly drop point, which corresponds to the happening of the Mott transition.

It is worth noticing that there exists a platform in the $\left\langle T_{z}^{2}\right\rangle$-$U$ curve
within the interaction region 3.55~eV$\leq U<$3.75~eV, indicating that the orbital polarization is especially
stable in the OSMP when $J_H=0.25U$.
Because the narrow $d_{3z^2-r^2}$ band with localized electron keeps half-filled in the OSMP,
the itinerant electrons in the wide $d_{x^2-y^2}$ can not transfer to the lower Hubbard subband of
the narrow-orbital, leading the orbital polarization to be fixed. Therefore, the orbital polarization
can also provide strong evidence for the occurrence of the OSMP.

\begin{figure}[htbp]
\includegraphics[scale=0.50]{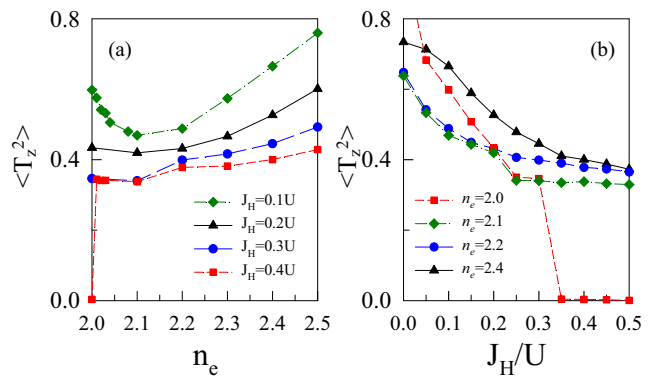}
\caption{(Color online) Orbital polarization in the hole-overdoped Ba$_2$CuO$_{4-\delta}$ when $U=3.6$ eV.
  (a) Local orbital squared moment as a function of $n_e$ for different Hund's rule couplings.
  (b) Comparing the $J_H$ dependencies of $\left\langle T_{z}^{2}\right\rangle$ for different $n_e$.
  }
\label{fig:Tz-x}
\end{figure}

The effect of the electron filling on the local orbital squared moments is
presented in Fig.~\ref{fig:Tz-x} for Ba$_2$CuO$_{4-\delta}$ when
$2.0\leq n_e\leq 2.5$. As shown in Fig.~\ref{fig:Tz-x}(a), the local
orbital squared moments slightly increase with the increasing
of electron filling when the system is in the metallic
phase for $n_e>2.2$. Near the half-filling region, the Hund's
rule coupling plays an essential role:
$\left\langle T_{z}^{2}\right\rangle$ decreases with increasing $n_e$ for
$J_H=0.1U$, because more electrons transfer to the low energy wide
$d_{x^2-y^2}$ band when the system has a large crystal-field splitting
and a small $J_H$.\cite{Oles-1993} When the Hund's rule coupling increases
to $J_H=0.3U$, an OSMT happens when $n_e\leq2.1$, and the squared moments
keep almost unchanged in the OSMP. This can also be seen in Fig.~\ref{fig:Tz-x}(b).

Totally speaking, the orbital polarization is suppressed by the
Hund's rule coupling in the metallic phase, but it becomes
almost constant in the OSMP, as one sees in Fig.~\ref{fig:Tz-x}(a) and \ref{fig:Tz-x}(b).
As a result, the orbital polarizations in the OSMP or near the edge
of the OSMP may be helpful for the occurrence of orbital-selective
superconductivity in Ba$_2$CuO$_{4-\delta}$  compound\cite{Scalapino-2019}.
Meanwhile, the orbital polarization in metallic phase
does not lead to orbital order due to the absence of lattice distortion.

\subsection{Discussion}

Thus far, we may expect that in the SC Ba$_2$CuO$_{4-\delta}$,
there exist two types of electrons: the narrow-band electrons near the edge of
the Mott localized state, and the wide-band electrons, resemble to an earlier two-band
hypothesis by Xiang {\it et al}. \cite{Xiang-2009}
Intuitively the multiorbital model for HTSC cuprates is more reasonable than the single-orbital model:
in conventional BCS superconductors, the ionic background and phononic vibrations provide the SC pairing
force field of Cooper pairs, and paired carriers are responsible for carrying the supercurrent;
in contrast, the carriers of the single-orbital $t-J$ model play duplicate roles, they not only create spin
 fluctuations to provide a SC pairing force but also carry a supercurrent.
Thus, the single-orbital $t-J$ model leads to a dilemma: the creation of the pairing force and carrying of the supercurrent
are competitive; the more the carriers participate in spin fluctuations, the less the carriers participate
in carrying the supercurrent, and {\it vice versa}.\cite{Bansil-2012,Graaf-2000,ZSArule-1985}
As a comparison, a multiorbital superconductor could avoid such a difficulty: the electrons in one or two
orbitals can contribute spin or orbital fluctuations, and electrons in another one or two orbitals
contribute SC
pairs and carry supercurrent.
In the same time, the multiorbital and OSMP characters of the compressed compound Ba$_2$CuO$_{4-\delta}$,
as well as of the compound Sr$_2$CuO$_{4-\delta}$,
imply that the spin fluctuations along with the orbital fluctuations may enhance the SC pairing
force and greatly lift $T_c$ in Ba$_2$CuO$_{4-\delta}$ and Sr$_2$CuO$_{4-\delta}$,
resembling to multiorbital
high-$T_c$  ironpnictide superconductors. Thus one could understand why
the SC critical temperatures of
Ba$_2$CuO$_{4-\delta}$ and Sr$_2$CuO$_{4-\delta}$ are about 70 and 90 K,
significantly larger than those of La$_2$CuO$_{4-\delta}$, which is about 30-40 K.
%

\section{conclusions}

In summary, we study the orbital selectivity of the  effective two-orbital Hubbard model
of Ba$_2$CuO$_{4-\delta}$ compound by using
the dynamical mean-field theory with  the {\it Lanczos} method as the impurity solver.
We demonstrate that Ba$_2$CuO$_{4-\delta}$ is an OSMP compound at half-filling
or is near the edge of the OSMP in the optimal hole-doping region,
and a stable orbital polarization can be observed in the OSMP regime.
These suggest that a local magnetic moment and spin or orbital fluctuations still exist,
and the OSMP and the orbital polarization
are significant features of  the hole-overdoped Ba$_2$CuO$_{4-\delta}$.
Our results are also applicable to Sr$_2$CuO$_{4-\delta}$ and other two-orbital cuprates.
Regarding the half-filled Ba$_2$CuO$_{3.5}$ as the parent phase,
our work provides a new perspective to understand the physics in the superconducting Ba$_2$CuO$_{4-\delta}$.

\section*{Acknowledgments}
 This work is supported by the National Natural Science Foundation of China (NSFC)
under the Grant Nos. 11474023, 11774350 and 11174036.

\bibliography{apssamp}

\end{document}